# Scattering of band carriers by neutral and charged impurities

**I.I. Boiko[1]**

Institute of Semiconductor Physics, NAS of Ukraine, prospect Nauky, 03028 Kiev, Ukraine

(Dated: November 13, 2010)

**Abstract.**

Mobility of band carriers scattered on donors, partially ionized, partially neutral, is considered in general and calculated for crystals of AIII-BV group. As neutral impurity we have constructed the hydrogen-like model, which gives well appointed short-range scattering potential. It is shown that dependence of mobility on temperature is determined by relation between number of ionized and neutral donors and by average energy of electrons.

## 1. Introduction

During long time investigations of scattering of band carriers by neutral impurities had no noticeable advance (see Refs. [1 – 5]). Several approaches to the problem of neutral impurity scattering were used, but wide recognition obtained only Erginsoy's formula of relaxation time for carriers momentum (see Ref.[1]):

$$\frac{1}{\tau} = \frac{20\,\varepsilon_L \hbar^3}{m^2 e^2} n_0 \ . \qquad (1)$$

Here $\varepsilon_L$ is dielectric constant of lattice, $m$ is effective mass, $n_0$ is concentration of scattering centers. But there are serious claims to method of scalar relaxation time in total [6, 7]. The attempts to improve agreement of theory with experimental data by the way of introducing in Eq. (1) some adjusting factor (see for instance [4] ) should be considered as very naive only.

Other direction of investigations is related to presentation of scattering potential for neutral impurity as rectangular spherically isotropic hole (see Ref. [8]). Limit case of this model is delta-shaped function in space (see Refs. [9, 10]). In this case there is no possibility to evaluate amplitude of interaction. There is also no way to derive rectangular or delta-shaped potential strictly as well-reasoned limit case of physically grounded interaction.

Bellow we consider mobility of band carriers, scattered by charged and neutral impurities; calculations will be based of quantum kinetic equation [9, 10]. For simplicity we use here only model of simple isotropic parabolic dispersion law for band carriers.

---

[1] E-mail: igorboiko@yandex.ru

As scattering system we consider shallow donors, which are partially ionized ; their extent of ionization depends on temperature. So in general case we have as neutral as charged scattering centers; relation between their concentration depends on temperature. We consider here only low temperatures and don't accept for calculation phonon scattering.

## 2. Scattering potential

### *2a. Delta-shaped potential*

The formulation "scattering of band electron on neutral point defect" is completely conditional, because Coulomb interaction of charged particle with really neutral point object is absent. Therefore neutral scattering center has to be some compact complex structure containing several different separated charges; and this structure is neutral only in total. For such case the range of forces is practically limited by geometrical size of complex center.

Very popular is simplest model of neutral scattering center, presented by delta-shaped potential of unknown amplitude:

$$\varphi_I(\vec{r}) = Y \delta(\vec{r}) \ . \tag{2}$$

Fourier component of this potential:

$$\varphi_I(\vec{q}) = \int_{-\infty}^{\infty} \exp(-i\vec{q}\vec{r}) \varphi_I(\vec{r}) \, d^3\vec{r} = Y \ . \tag{3}$$

Note, that the value $\varphi_I(\vec{q})$ does not depend on wave-vector $\vec{q}$.

### *2b. Charged impurity*

Fourier component of Coulomb potential generated by solitary charged impurity has the form (see Refs. [9, 10])

$$\varphi_{CI}(\vec{q},\omega) = 4\pi e_I \, \delta(\omega) / q^2 \varepsilon_L \ . \tag{4}$$

Taking in account screening of scattering potential by band carriers one obtains:

$$\varphi_{CI}(\vec{q},\omega) = \frac{4\pi e_I}{q^2[\varepsilon_L + \Delta\varepsilon(\vec{q},0)]} \delta(\omega) \ . \tag{5}$$

Here $\varepsilon_L$ is dielectric constant of lattice, $\Delta\varepsilon(\vec{q},\omega)$ is dielectric function for band carriers. Farther we use the last in the form

$$\Delta \varepsilon\ (0,\vec{q}) = \frac{e^2 \sqrt{8m^3 k_B T}}{q^2 \sqrt{\pi} \varepsilon_L \hbar^3} F_{-1/2}(\varepsilon_F / k_B T)\ ; \tag{6}$$

Here $F_{-1/2}(\eta)$ is Fermi-integral (see Ref. [8]):

$$F_{-1/2}(\eta) = \frac{2}{\sqrt{\pi}} \int \frac{t^{1/1}}{1+\exp(t-\eta)}\, dt\ . \tag{7}$$

It is convenient to write the expression (6) in the form

$$\Delta \varepsilon\ (0,\vec{q}) = q_0^2 \varepsilon_L^2 / q^2 . \tag{8}$$

Then we obtain

$$\varphi_{CI}(\vec{q},\omega) = \frac{4\pi e_I}{\varepsilon_L (q^2 + q_0^2)} \delta(\omega)\ . \tag{9}$$

Comparison of the forms (3) and (8) shows that even high-screened Coulomb potential cannot imitate the delta-shaped potential. The reason of this result is evident: the screening cuts Coulomb interaction at long distances and is not important for short distance interaction.

Correlator of screened potentials:

$$\langle \varphi_{CI}^2 \rangle_{\omega,\vec{q}} = \langle \varphi_{CI}^2 \rangle_q \delta(\omega) = \frac{32\pi^3 e^2 N_{CI}}{\varepsilon_L^2 (q^2 + q_0^2)^2} \delta(\omega) . \tag{10}$$

*2c. Hydrogen-shape neutral impurity*

Consider donor impurity which has the structure, similar to spherically symmetrical hydrogen atom. Space density of negative charge $\chi$ can be presented by following relation:

$$\chi^{(-)}(\rho) = -e|\psi(\rho)|^2 . \tag{11}$$

Here $\psi(\rho) \sim \exp(-\rho/r_B)$ is wave function of electron of shallow donor; the value

$$r_B = \hbar^2 \varepsilon_L / me^2 \tag{12}$$

is Bohr-radius of exterior donor electron; $m$ is effective mass.. The density of charge $\chi(\rho)$ is normalized by the relation

$$4\pi \int_0^\infty \chi^{(-)}(\rho) \rho^2 d\rho = -e . \tag{13}$$

Electrostatic potential of the positive kernel of impurity atom in crystal is

$$\varphi^+(r) = \frac{e}{\varepsilon_L r}.  \quad (14)$$

Total scattering potential of neutral center, generated by distributed negative charge of exterior donor electron and point positive charge in center, is

$$\varphi_{NI}(r) = \varphi^+(r) + \varphi^-(r) = \frac{e}{\varepsilon_L r}\left[1 + \frac{r}{r_B}\right]\exp(-2r/r_B). \quad (15)$$

Several examples for the space distribution of potential $\varphi_{NI}(r)$ are presented on Fig. 1. Here $K(r,\gamma) = r_0 \varepsilon_L \varphi_{NI}(r)/e$ and $\gamma = r_0/r_B$. The value of radius $r_B$ determines range of action for scattering center, and $r_0$ is scale-factor.

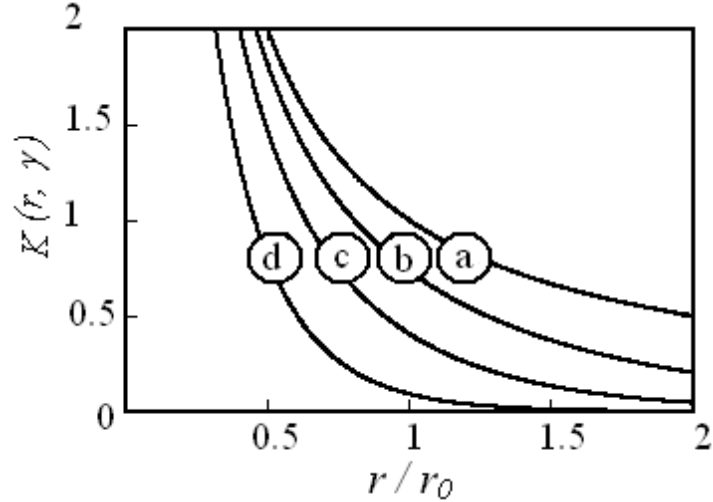

Fig. 1.  a) $\gamma = 0$; b) $\gamma = 1$; c) $\gamma = 2$; d) $\gamma = 4$.

Fourier component of potential (15) has the following form:

$$\varphi_{NI}(\vec{q},\omega) = \frac{4\pi}{\varepsilon_L}\frac{e(q^2 + 2q_B^2)}{(q^2 + q_B^2)^2}\delta(\omega). \quad (16)$$

Here $q_B = 2/r_B = 2me^2/\varepsilon_L \hbar^2$.

One can see that wave-vector $q_B$ is natural measure for distribution in $q$-space of Fourier-component for scattering potential, generated by neutral impurity. For $n$-GaAs we have: $m = 0.067\,m_0$ and $\varepsilon_L = 12.5$; therefore we obtain: $q_B^2 = 4.108 \cdot 10^{12}\,cm^{-2}$. It follows from here that noticeable screening of the short-range potential (16) by band electrons arrives in mentioned crystal at concentrations $n > 10^{17} cm^{-3}$. Assuming in the expression (16) formally $q_B = 0$ (that is $r_B \to \infty$), we obtain the Coulomb form (4). Assuming there $q/q_B \to 0$, we obtain the model form (3), used for short range scattering centres.

From Eq. (16) follows the correlator of scattering potentials for neutral centres:

$$\langle \varphi_{NI}^2 \rangle_{\omega,\vec{q}} = \langle \varphi_{NI}^2 \rangle_q \delta(\omega) = \frac{32\pi^3 e^2 N_{NI} (q^2 + 2q_B^2)^2}{\varepsilon_L^2 (q^2 + q_B^2)^4} \delta(\omega). \qquad (17)$$

Here $N_{NI}$ is concentration of neutral impurities.

Due to short range of considered scattering center there is no need to involve in consideration screening of scattering potential by band electrons.

### 3. Mobility of band carriers

Consider impurity system as donors, partially ionized, partially neutral. The degree of ionization depends on temperature $T$. Write the relation between densities of ionized and neutral impurities as

$$N_D = N_{DN} + N_{DI} = N_{NI} + N_{CI}. \qquad (18)$$

Below we assume that band electrons concentration $n$ equals to concentration of ionized donors

$$n = N_{DI} = N_{CI} = N_D \{1 + \exp[(\varepsilon_F - \varepsilon_D)/k_B T]\}^{-1}. \qquad (19)$$

Here $\varepsilon_F$ is Fermi energy, $\varepsilon_D < 0$ presents energy level for donors.

Calculate mobility $\mu$ of band carriers using the formulae (see Ref. [11])

$$\beta = \mu^{-1} = \beta_{CI} + \beta_{NI}; \qquad (20)$$

$$\beta_{CI} = \frac{em^2}{24\pi^4 \hbar^3 n} \int_0^\infty \frac{q^3 \, dq}{1 + \exp(\hbar^2 q^2 / 8m\, k_B T - \eta)} \langle \varphi_{CI}^2 \rangle_q; \qquad (21)$$

$$\beta_{NI} = \frac{em^2}{24\pi^4 \hbar^3 n} \int_0^\infty \frac{q^3 \, dq}{1 + \exp(\hbar^2 q^2 / 8m\, k_B T - \eta)} \langle \varphi_{NI}^2 \rangle_q. \qquad (22)$$

The value $\beta_{CI}$ presents deposit of charged impurities in (reverse) mobility of band carriers, the value $\beta_{NI}$ relates to neutral impurities subsequently.

The following numerical calculations we carry out for set of AIII-BV crystals (see Ref. [11] and Table 1):

TABLE – 1

|   | AIII-BV | $m/m_0$ | $\varepsilon_L$ | $-\varepsilon_D$ (eV) |
|---|---------|---------|-----------------|------------------------|
| 1 | GaAs    | 0.067   | 12.5            | 0.008                  |
| 2 | GaSb    | 0.05    | 15              | 0.003                  |
| 3 | InP     | 0.07    | 14              | 0.008                  |
| 4 | InSb    | 0.013   | 17              | 0.0007                 |
| 5 | InAs    | 0.02    | 14              | 0.002                  |

Results of calculations of mobility, based on formulae (21), (22), are shown on Fig. 2 (a – g). Here mobility $\mu$ is presented in CGSE units. Numbers on curves correspond to numbers in Table 1.

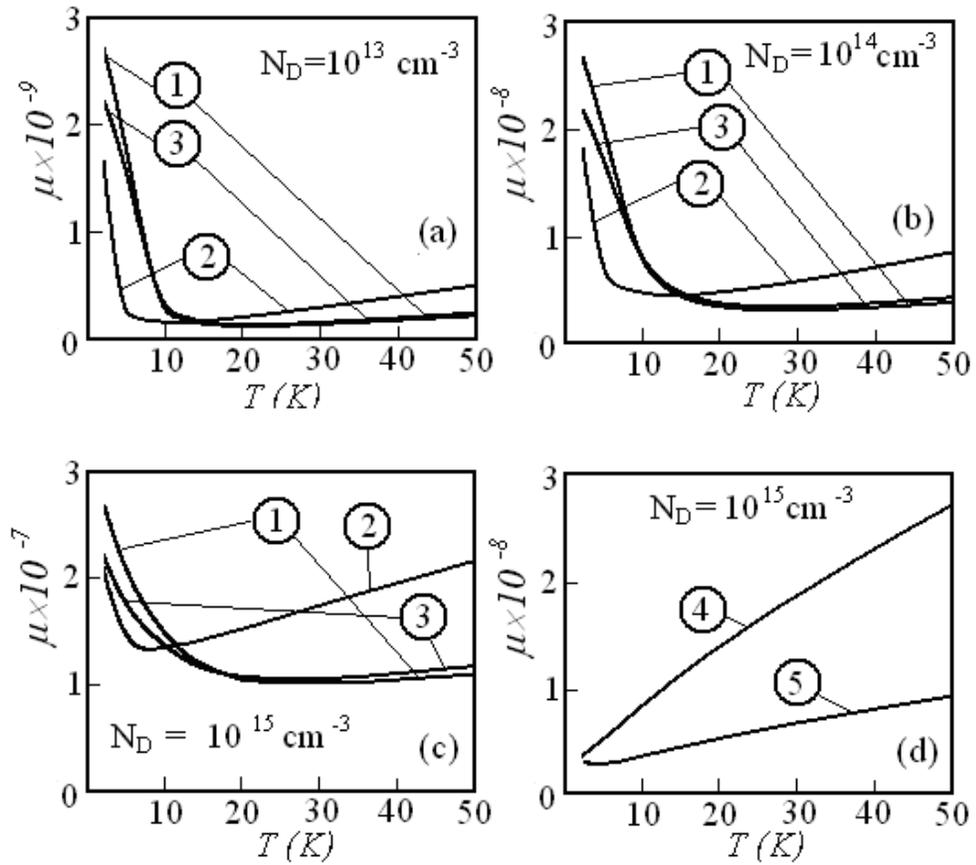

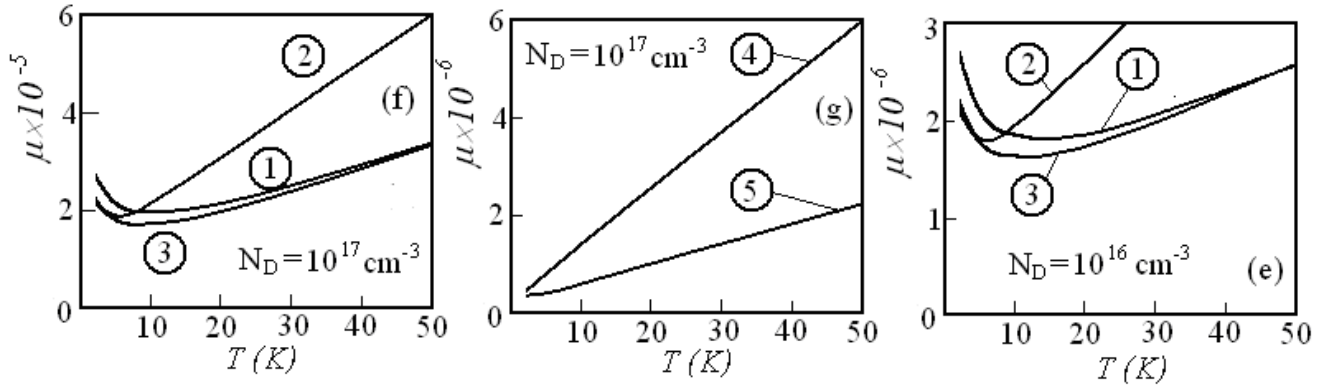

Fig. 2.

Temperature dependence of mobility $\mu$ for considered crystals appears as result of competition of two processes. Change of temperature changes as the number of ionized centres as average energy of electronsl. First process dominates at lower temperatures, second – at more high temperatures. Therefore, calculated dependence of dimensionless mobility $M$ on temperature $T$ becomes non-monotonous.

For comparison show here Fig. 3 which reproduces figure 4.5(b) from Ref. [8]. That figure relates to scattering of electrons by "neutral" impurities. Here curve-1 is constructed on the base of Erginsoy's theory [1], curves-2 and 3 – on the base of theoretical calculations of N. Sclar [11] and T. McGill with R.Baron [12] consequently. Our curves shown on Fig. 4 have at low temperatures the same stile as curves-2 and 3 on Fig. 3. This comparison supports our critical relation to Erginsoy's formula.

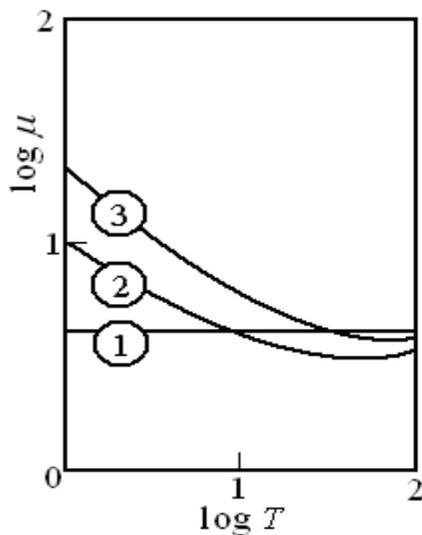 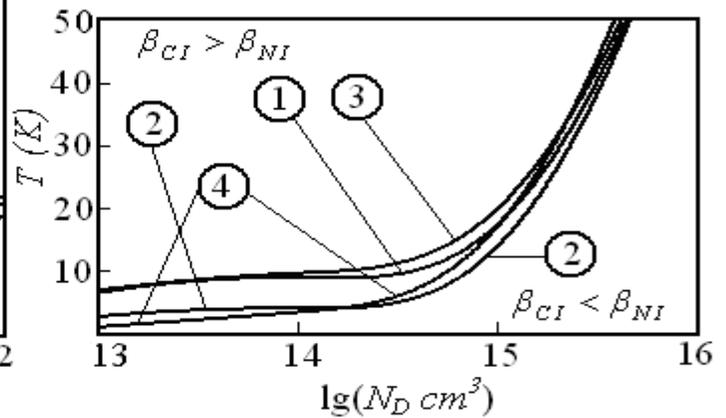

Fig. 3.                     Fig. 4 .

## 4. Discussion

To compare contribution of neutral and charged impurities to mobility introduce border temperature $T^*$ by the relation $\beta_{CI}(T^*) = \beta_{NI}(T^*)$. Connection between temperature $T^*$ and donor concentration $N_D$ is presented on Fig. 4 by five lines (corresponding to five different crystals). These lines divide the plane $\{N_D, T\}$ on two areas. In top area (there $T > T^*$) scattering on charged donor prevails; in lower area (there $T < T^*$) scattering on neutral impurities dominates. The numbers of curves correspond to that in Table-1.

Now compare results of this article with results which can be obtained on the base of calculations carried out on the base of traditional $\tau$-approximation (see Ref. [5]). Result of comparison is shown on Fig. 5 (Here B-lines relate to the calculations of this article, A-lines are constructed with the help of corresponding formulae represented in monograph of Anselm (see Ref. [5]). One can see that divergence is quite noticeable.

Fig. 6 represents our theoretical curve (solid line) and experimental curve (dashed line) obtained for InSb by H.J. Hrostowski *et al.* (see Refs. [13, 14] ). Due to limited accuracy of parameters, presented in Table 1, one can consider concordance of these lines quite satisfactory.

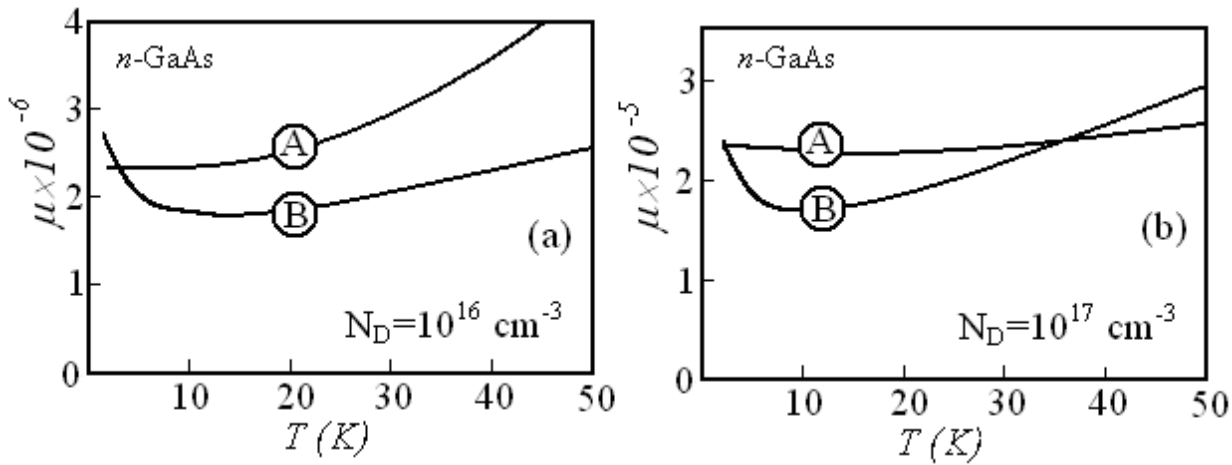

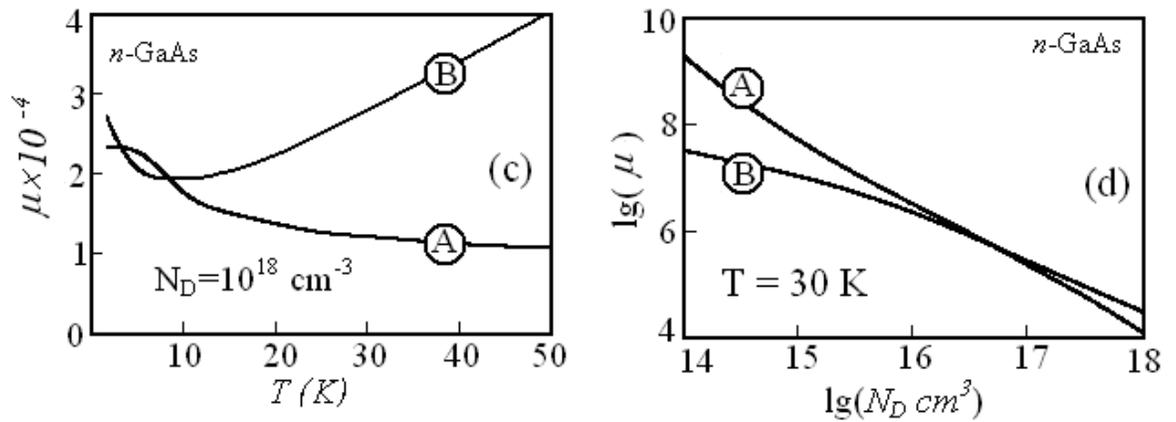

Fig. 5.

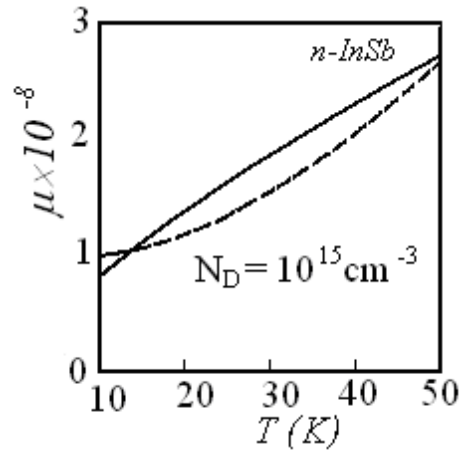

Fig. 6.

---